\documentclass{elsart}

\usepackage{graphics}
\usepackage{graphicx,float}
\usepackage{epsfig}
\usepackage{amsmath,amsfonts,amsgen}
\usepackage{dcolumn}
\usepackage[all]{xy}
\usepackage{makeidx,color}
\usepackage{color}

\newcolumntype{d}[1]{D{.}{\cdot}{#1}}
\newcolumntype{.}{D{.}{.}{-1}}

\newcommand\tabcaption{\def\@captype{table}\caption}
\newcommand{\be}{\begin{eqnarray*}}
\newcommand{\ee}{\end{eqnarray*}}
\newcommand{\ibe}{\begin{eqnarray}}
\newcommand{\iee}{\end{eqnarray}}

\def\3hf{\frac{3}{2}}

\def\np1{n+1}

\def\ip3hf{i+\frac{3}{2}}
\def\jp3hf{j+\frac{3}{2}}

\begin{document}
\begin{frontmatter}

\title{Fractional derivatives of constant and variable orders applied to anomalous
relaxation models in heat-transfer problems}
\author{Xiao-Jun Yang$^{1,2,\ast}$}
\corauth[cor1]{Corresponding author: Tel.: 086-67867615; E-Mail: dyangxiaojun@163.com (X. J. Yang)}

\address{$^{1}$ School of Mechanics and Civil Engineering, China University of Mining and Technology, Xuzhou 221116, China}
\address{$^{2}$ State Key Laboratory for Geomechanics and Deep Underground Engineering,
China University of Mining and Technology, Xuzhou 221116, China}
\begin{abstract}
In the present paper, we address a class of the fractional derivatives of constant
and variable orders for the first time. Fractional-order relaxation
equations of constants and variable orders in the sense of Caputo type are
modeled from mathematical view of point. The comparative results of the
anomalous relaxation among the various fractional derivatives are also
given. They are very efficient in description of the complex phenomenon
arising in heat transfer.
\end{abstract}


\end{frontmatter}

\section{Introduction}

Fractional derivatives (FDs) were utilized to model the complex phenomenon
in science and engineering practice \cite{1,2,3,4,5,6,7}. FDs of constant order, e.g.,
RiemannLiouville and Caputo FDs were considered to describe the anomalous
relaxation \cite{2,5,8,9,10}. The Riemann-Liouville type FD of variable-order was
proposed in \cite{11,12} and its extended versions were considered in \cite{13}. The
Caputo type FD of variable-order was developed in \cite{14} and its extended
versions were reported in \cite{15,16}.

Recently, FDs involving the kernels of the exponential and Mittag-Leffler
functions were reported in \cite{17,18,19,20,21,22,23,24}. For example, the FD with respect to the
kernel of the exponential function in the sense of the Caputo type was
firstly reported in \cite{17,18,19,20} and the FD with the aid of the kernel of the
exponential function in the sense of the Riemann-Liouville type was proposed
in \cite{21} and developed in \cite{22}. The FD with the help of the kernel of the
stretched exponential function in the sense of the Caputo type was presented
in \cite{23}. The FDs containing the Mittag-Leffler function kernels in the sense
of the Caputo and Riemann-Liouville types were developed in \cite{24}.

Due to the above FDs in the different kernels, a class of the fractional
derivatives of constant and variable orders, such as has not been reported.
Motivated by the above ideas, the brief aims of the previous paper are to
suggest the FD of constant order with respect to the Mittag-Leffler function
and stretched Mittag-Leffler and exponential functions kernels in the sense
of Gaussian-like type, the variableorder FDs in the Caputo-Fabrizio,
Sun-Hao-Zhang- Baleanu and Atangana-Baleanu types as well as Mittag-Leffler
and exponential functions and their stretched and generalized versions and
to present their applications for handling the anomalous relaxation models
in heat-transfer problems.

The paper is structured as follows. In Section 2, we present the concepts of
the FDs of constant and variable orders. In Section 3, the models for the
anomalous relaxation arising in heat-transfer problems are considered.
Finally, the conclusion is outlined in Section 4.

\section{FDs of constant and variable orders}

Let $\Xi \left( \mu \right)$ be the differentiable function defined on the
interval $\left[ {a,b} \right]$, as well as $\Im \left( \omega \right)$ and
$\Im \left( {\omega \left( x \right)} \right)_{ }$ be normalization
constants with respect to the orders $\omega $ $(0\le \omega \le1)$ and $\omega \left( x \right)$ $(0\le \omega \left( x \right) \le1)$ ,
respectively. For more details of the normalization constants, readers refer
to \cite{17,18,19,20,21,22,23,24}.

\subsection{FDs of constant order}

FD of the Caputo type with respect to the singular power-law kernel is given
by \cite{2,5,10}:
\begin{equation}
\label{eq1}
D_{a^+}^{\left( \omega \right)} \Xi \left( x \right)=\frac{1}{\Gamma \left(
{1-\omega } \right)}\int\limits_a^x {\frac{1}{\left( {x-\mu } \right)^\omega
}\left( {\frac{d\Xi \left( \mu \right)}{d\mu }} \right)d\mu },
\end{equation}
where $a\le x$, $n$ is an integer, and $\Gamma \left( \cdot \right)$ denotes
the Gamma function.

FD of the Caputo-Fabrizio type with respect to the exponential kernel,
denoted by \cite{17,18,19}
\begin{equation}
\label{eq2}
y=\phi _1 \left( {x,\omega } \right)=\exp \left( {-\frac{\omega }{1-\omega
}x} \right),
\end{equation}
is given by:
\begin{equation}
\label{eq3}
{ }^{CF}D_x^{\left( \omega \right)} \Xi \left( x \right)=\frac{\left(
{2-\omega } \right)\Im \left( \omega \right)}{2\left( {1-\omega }
\right)}\int\limits_a^x {\exp \left( {-\frac{\omega }{1-\omega }\left(
{x-\mu } \right)} \right)\Xi ^{\left( 1 \right)}\left( \mu \right)d\mu } ,
\end{equation}
where $a\le x$.

For $\Im \left( \omega \right)=2/\left( {2-\omega } \right)$, Eq.(\ref{eq3}) is
given by \cite{19}:
\begin{equation}
\label{eq4}
{ }_\ast ^{LNCF} D_x^{\left( \omega \right)} \Xi \left( x
\right)=\frac{1}{1-\omega }\int\limits_0^x {\exp \left( {-\frac{\omega
}{1-\omega }\left( {x-\mu } \right)} \right)\Xi ^{\left( 1 \right)}\left(
\mu \right)d\mu } .
\end{equation}
FD of the Sun-Hao-Zhang-Baleanu type with respect to the stretched
exponential function, denoted by
\begin{equation}
\label{eq5}
y = \phi _{\rm{2}} \left( {x,\omega } \right) = \exp \left( { - \frac{\omega }{{1 - \omega }}x^\omega  } \right),
\end{equation}
is given by \cite{23}:
\begin{equation}
\label{eq6}
{ }^{SHZD}D_x^{\left( \omega \right)} \Xi \left( x \right)=\frac{\Im \left(
\omega \right)}{\left( {1-\omega } \right)^{1/\omega }}\int\limits_a^x {\exp
\left( {-\frac{\omega }{1-\omega }\left( {x-\mu } \right)^\omega }
\right)\Xi ^{\left( 1 \right)}\left( \mu \right)d\mu } ,
\end{equation}
where $a\le x$.

For simplicity, we have \cite{23}
\begin{equation}
\label{eq7}
\Im \left( \omega \right)=\Gamma \left( {\omega +1} \right)
\end{equation}
such that
\begin{equation}
\label{eq8}
{ }^{SHZD}D_x^{\left( \omega \right)} \Xi \left( x \right)=\frac{\Gamma
\left( {1+\omega } \right)}{\left( {1-\omega } \right)^{1/\omega
}}\int\limits_0^x {\exp \left( {-\frac{\omega }{1-\omega }\left( {x-\mu }
\right)^\omega } \right)\Xi ^{\left( 1 \right)}\left( \mu \right)d\mu } .
\end{equation}
FD with respect to the Mittag-Leffler function kernel, denoted by
\begin{equation}
\label{eq9}
y = \phi _{\rm{3}} \left( {x,\omega } \right) = {\rm E}_\omega  \left( { - \frac{\omega }{{1 - \omega }}x} \right),
\end{equation}
is defined by:
\begin{equation}
\label{eq10}
{ }^{YC}D_x^{\left( \omega \right)} \Xi \left( x \right)=\frac{\Im \left(
\omega \right)}{1-\omega }\int\limits_a^x {{\rm E}_\omega \left(
{-\frac{\omega }{1-\omega }\left( {x-\mu } \right)} \right)\Xi ^{\left( 1
\right)}\left( \mu \right)d\mu } ,
\end{equation}
where $a\le x$.

FD of the Atangana-Baleanu type with respect to the stretched Mittag-Leffler
function kernel, denoted by
\begin{equation}
\label{eq11}
y = \phi _{\rm{4}} \left( {x,\omega } \right) = {\rm E}_\omega  \left( { - \frac{\omega }{{1 - \omega }}x^\omega  } \right),
\end{equation}
is given by \cite{24}:
\begin{equation}
\label{eq12}
{ }^{ABC}D_x^{\left( \omega \right)} \Xi \left( x \right)=\frac{\Im \left(
\omega \right)}{1-\omega }\int\limits_a^x {{\rm E}_\omega \left(
{-\frac{\omega }{1-\omega }\left( {x-\mu } \right)^\omega } \right)\Xi
^{\left( 1 \right)}\left( \mu \right)d\mu } ,
\end{equation}
where $a\le x$.

FD of the Caputo-Fabrizio type with respect to the Gaussian-function kernel,
denoted by
\begin{equation}
\label{eq13}
y = \phi _{\rm{5}} \left( {x,\omega } \right) = \exp \left( {-\frac{\omega }{1-\omega }x^2} \right)=\sum\limits_{i=0}^\infty
{\left( {-\frac{\omega }{1-\omega }} \right)^i\frac{x^{2i}}{\Gamma \left(
{1+i} \right)}},
\end{equation}
is given by \cite{20}:
\begin{equation}
\label{eq14}
{ }^{CF}D_x^{\left( \omega \right)} \Xi \left( x \right)=\frac{1+\omega
^2}{\sqrt {\pi ^\omega \left( {1-\omega } \right)} }\int\limits_a^x {\exp
\left( {-\frac{\omega }{1-\omega }\left( {x-\mu } \right)^2} \right)\Xi
^{\left( 1 \right)}\left( \mu \right)d\mu } ,
\end{equation}
where $a\le x$.

FD with respect to the stretched exponential function kernel in the sense of
Gaussian-like type, denoted by
\begin{equation}
\label{eq15}
y = \phi _{\rm{6}} \left( {x,\omega } \right) = \exp \left( { - \frac{\omega }{{1 - \omega }}x^{2\omega } } \right),
\end{equation}
is defined by:
\begin{equation}
\label{eq16}
{ }^{GYC}D_x^{\left( \omega \right)} \Xi \left( x \right)=\frac{1+\omega
^2}{\sqrt {\pi ^\omega \left( {1-\omega } \right)} }\int\limits_a^x {\exp
\left( {-\frac{\omega }{1-\omega }\left( {x-\mu } \right)^{2\omega }}
\right)\Xi ^{\left( 1 \right)}\left( \mu \right)d\mu } ,
\end{equation}
where $a\le x$.

FD with respect to the stretched Mittag-Leffler function kernel in the sense
of Gaussian-like type, denoted by
\begin{equation}
\label{eq17}
y = \phi _{\rm{7}} \left( {x,\omega } \right) = {\rm E}_\omega  \left( { - \frac{\omega }{{1 - \omega }}x^{2\omega } } \right),
\end{equation}
is defined by:
\begin{equation}
\label{eq18}
{ }^{GYGC}D_x^{\left( \omega \right)} \Xi \left( x \right)=\frac{1+\omega
^2}{\sqrt {\pi ^\omega \left( {1-\omega } \right)} }\int\limits_a^x {{\rm
E}_\omega \left( {-\frac{\omega }{1-\omega }\left( {x-\mu } \right)^{2\omega
}} \right)\Xi ^{\left( 1 \right)}\left( \mu \right)d\mu } ,
\end{equation}
where $a\le x$.

The LTs of the functions $x^{-1-\omega }\mbox{/}\Gamma \left( {-\omega }
\right)$ and ${\rm E}_\omega \left( {-x^\omega } \right)$ are given by (see
\cite{2,12}):
\begin{equation}
\label{eq19}
\widehat{L}\left[ {\frac{x^{-1-\omega }}{\Gamma \left( {-\omega } \right)}}
\right]=s^\omega ,
\end{equation}
\begin{equation}
\label{eq20}
\widehat{L}\left[ {{\rm E}_\omega \left( {-\beta x^\omega } \right)}
\right]=\frac{s^{\omega -1}}{s^\omega +\beta },
\end{equation}
respectively, where $\widehat{L}$ is the LT operator with respect to $x$,
and $\beta $ is a constant.

The LTs of the FDs are as follows:
\begin{equation}
\label{eq21}
\widehat{L}\left[ {{ }^{SHZD}D_x^{\left( \omega \right)} \Xi \left( x
\right)} \right]=\frac{{\Im \left( \omega  \right)}}{{\left( {1 - \omega } \right)^{1/\omega } }}\sum\limits_{i = 0}^\infty  {\left( { - \frac{\omega }{{1 - \omega }}} \right)^i \frac{{\Gamma \left( {1 + i\omega } \right)}}{{\Gamma \left( {1 + i} \right)}}\frac{{\left[ {s\hat \Xi \left( s \right) - \Xi \left( 0 \right)} \right]}}{{s^{i\omega  + 1} }}},
\end{equation}
\begin{equation}
\label{eq22}
\widehat{L}\left[ {{ }^{YC}D_x^{\left( \omega \right)} \Xi \left( x \right)}
\right]=\frac{{\Im \left( \omega  \right)}}{{1 - \omega }}\sum\limits_{i = 0}^\infty  {\left( { - \frac{\omega }{{1 - \omega }}} \right)^i \frac{{\Gamma \left( {i + 1} \right)}}{{\Gamma \left( {i\omega  + 1} \right)}}\frac{{\left[ {s\hat \Xi \left( s \right) - \Xi \left( 0 \right)} \right]}}{{s^{i + 1} }}},
\end{equation}
\begin{equation}
\label{eq23}
\widehat{L}\left[ {{ }^{GYC}D_x^{\left( \omega \right)} \Xi \left( x
\right)} \right]=\frac{{1 + \omega ^2 }}{{\sqrt {\pi ^\omega  \left( {1 - \omega } \right)} }}\sum\limits_{i = 0}^\infty  {\left( { - \frac{\omega }{{1 - \omega }}} \right)^i \frac{{\Gamma \left( {1 + 2\omega i} \right)}}{{\Gamma \left( {1 + i} \right)}}\frac{{\left[ {s\hat \Xi \left( s \right) - \Xi \left( 0 \right)} \right]}}{{s^{2i\omega  + 1} }}},
\end{equation}
\begin{equation}
\label{eq24}
\widehat{L}\left[ {{ }^{GYGC}D_x^{\left( \omega \right)} \Xi \left( x
\right)} \right]=\frac{{1 + \omega ^2 }}{{\sqrt {\pi ^\omega  \left( {1 - \omega } \right)} }}\sum\limits_{i = 0}^\infty  {\left( { - \frac{\omega }{{1 - \omega }}} \right)^i \frac{{\left[ {s\hat \Xi \left( s \right) - \Xi \left( 0 \right)} \right]}}{{\Gamma \left( {1 + i\omega } \right)s^{2i\omega  + 1} }}}.
\end{equation}
For the details of the LTs of Eqs.(1,3,12), see \cite{2,17,24}.

A plot of the different kernels of the FDs for $\omega =0.6$ is displayed in
Figure 1.

\hspace{3mm}
\begin{center}
\includegraphics[width=5in]{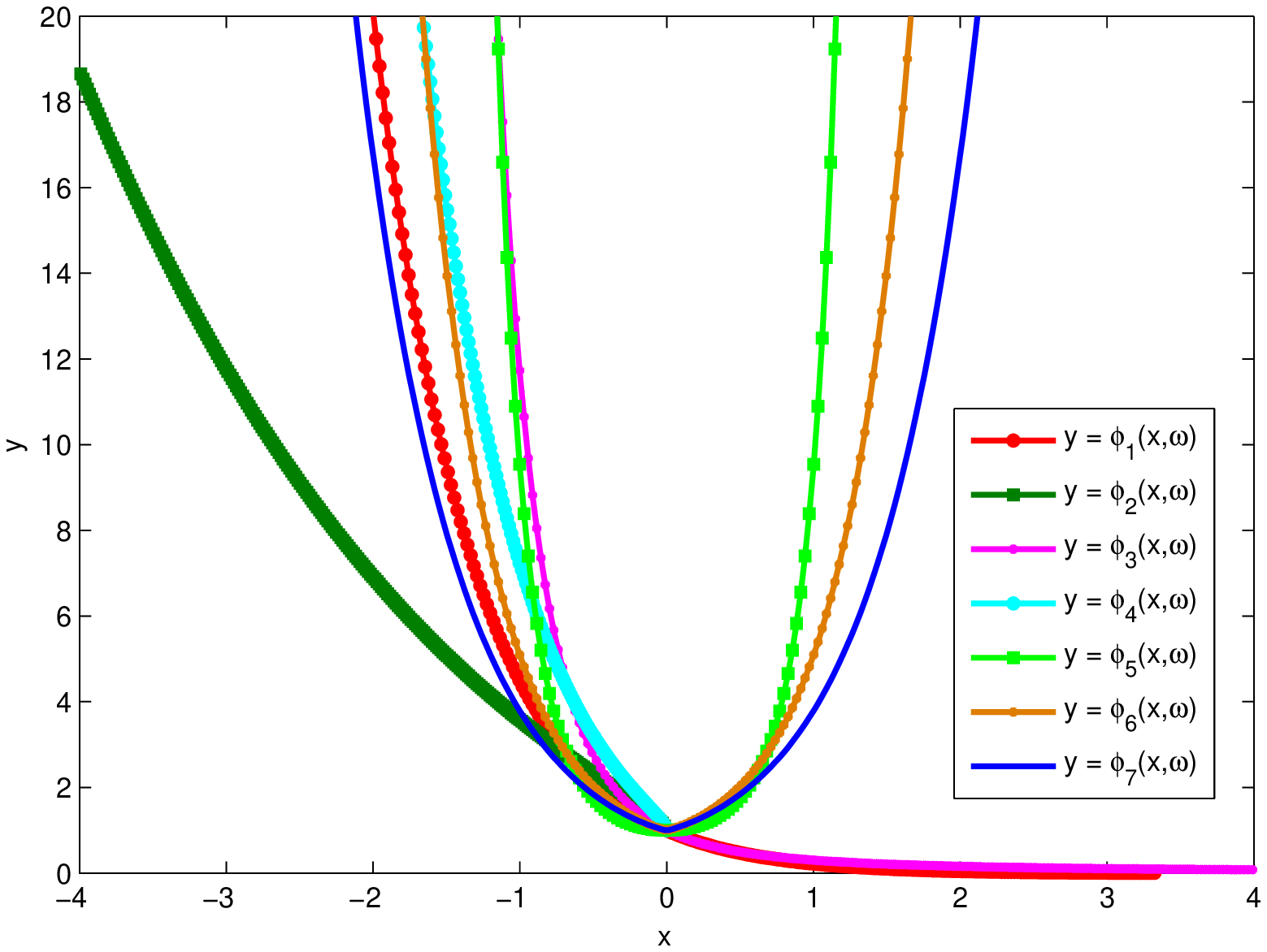}\\
{\footnotesize Figure 1. The plot of the different kernels of the FDs for$\omega =0.6$.}
\end{center}
\label{fig2}
\hspace{3mm}

\subsection{FDs of variable order}

Variable-order FD of Caputo type in term of the singular power-law kernel is
given by \cite{15,16}:
\begin{equation}
\label{eq25}
D_{a^+}^{\left( {\omega \left( x \right)} \right)} \Xi \left( x
\right)=\frac{1}{\Gamma \left( {1-\omega \left( x \right)}
\right)}\int\limits_a^x {\frac{1}{\left( {x-\mu } \right)^{\omega \left( x
\right)}}\left( {\frac{d\Xi \left( \mu \right)}{d\mu }} \right)d\mu },
\end{equation}
where $a\le x$ and $n$ is an integer.

Variable-order FD of Caputo-Fabrizio type in term of the exponential kernel,
denoted by
\begin{equation}
\label{eq26}
y=\phi _1 \left( {x,\omega \left( t \right)} \right)=\exp \left(
{-\frac{\omega \left( x \right)}{1-\omega \left( x \right)}x}
\right),
\end{equation}
is defined by:
\begin{equation}
\label{eq27}
{ }^{VYCF}D_x^{\left( {\omega \left( x \right)} \right)} \Xi \left( x
\right)=\frac{1}{1-\omega \left( x \right)}\int\limits_a^x {\exp \left(
{-\frac{\omega \left( x \right)}{1-\omega \left( x \right)}\left( {x-\mu }
\right)} \right)\Xi ^{\left( 1 \right)}\left( \mu \right)d\mu } ,
\end{equation}
where $a\le x$.

Variable-order FD of the stretched exponential function in the sense of the
Sun-Hao-Zhang- Baleanu type, denoted by
\begin{equation}
\label{eq28}
y=\phi _2 \left( {x,\omega \left( t \right)} \right)=\exp \left(
{-\frac{\omega \left( x \right)}{1-\omega \left( x \right)}x^{\omega \left(
x \right)}} \right),
\end{equation}
is defined by:
\begin{equation}
\label{eq29}
{ }^{VYSHZD}D_x^{\left( \omega \right)} \Xi \left( x \right)=\frac{\Im
\left( {\omega \left( x \right)} \right)}{1-\omega \left( x
\right)}\int\limits_a^x {\exp \left( {-\frac{\omega \left( x
\right)}{1-\omega \left( x \right)}\left( {x-\mu } \right)^{\omega \left( x
\right)}} \right)\Xi ^{\left( 1 \right)}\left( \mu \right)d\mu } ,
\end{equation}
where $a\le x$.

Generalized variable-order FD of the stretched exponential function, denoted
by
\begin{equation}
\label{eq30}
y=\phi _3 \left( {x,\omega \left( t \right)} \right)=\exp \left( {-x^{\omega
\left( x \right)}} \right),
\end{equation}
is defined by:
\begin{equation}
\label{eq31}
{ }^{GVYSHZD}D_x^{\left( \omega \right)} \Xi \left( x \right)=\frac{\Im
\left( {\omega \left( x \right)} \right)}{1-\omega \left( x
\right)}\int\limits_a^x {\exp \left( {-\left( {x-\mu } \right)^{\omega
\left( x \right)}} \right)\Xi ^{\left( 1 \right)}\left( \mu \right)d\mu } ,
\end{equation}
where $a\le x$.

Variable-order FD involving the Mittag-Leffler function kernel, denoted by
\begin{equation}
\label{eq32}
y=\phi _4 \left( {x,\omega \left( t \right)} \right)={\rm E}_{\omega \left(
x \right)} \left( {-\frac{\omega \left( x \right)}{1-\omega \left( x
\right)}x} \right),
\end{equation}
is given by:
\begin{equation}
\label{eq33}
{ }^{VYGC}D_x^{\left( {\omega \left( x \right)} \right)} \Xi \left( x
\right)=\frac{1}{1-\omega \left( x \right)}\int\limits_a^x {{\rm E}_{\omega
\left( x \right)} \left( {-\frac{\omega \left( x \right)}{1-\omega \left( x
\right)}\left( {x-\mu } \right)} \right)\Xi ^{\left( 1 \right)}\left( \mu
\right)d\mu } .
\end{equation}
Generalized variable-order FD involving the Mittag-Leffler function kernel,
denoted by
\begin{equation}
\label{eq34}
y=\phi _5 \left( {x,\omega \left( t \right)} \right)={\rm E}_{\omega \left(
x \right)} \left( {-x} \right),
\end{equation}
is given by:
\begin{equation}
\label{eq35}
{ }^{GVYGC}D_x^{\left( {\omega \left( x \right)} \right)} \Xi \left( x
\right)=\frac{1}{1-\omega \left( x \right)}\int\limits_a^x {{\rm E}_{\omega
\left( x \right)} \left( {-\left( {x-\mu } \right)} \right)\Xi ^{\left( 1
\right)}\left( \mu \right)d\mu } .
\end{equation}
The Atangana-Baleanu-type variable-order FD involving the stretched
Mittag-Leffler function kernel, denoted by
\begin{equation}
\label{eq36}
y=\phi _6 \left( {x,\omega \left( t \right)} \right)={\rm E}_{\omega \left(
x \right)} \left( {-\frac{\omega \left( x \right)}{1-\omega \left( x
\right)}x^{\omega \left( x \right)}} \right),
\end{equation}
is defined by:
\begin{equation}
\label{eq37}
{ }^{VYABC}D_x^{\left( {\omega \left( x \right)} \right)} \Xi \left( x
\right)=\frac{1}{\Gamma \left( {1-\omega \left( x \right)}
\right)}\int\limits_a^x {{\rm E}_{\omega \left( x \right)} \left(
{-\frac{\omega \left( x \right)}{1-\omega \left( x \right)}\left( {x-\mu }
\right)^{\omega \left( x \right)}} \right)\Xi ^{\left( 1 \right)}\left( \mu
\right)d\mu } ,
\end{equation}
where $a\le x$.

Generalized variable-order FD involving the stretched Mittag-Leffler
function kernel, denoted by
\begin{equation}
\label{eq38}
y=\phi _7 \left( {x,\omega \left( t \right)} \right)={\rm E}_{\omega \left(
x \right)} \left( {-x^{\omega \left( x \right)}}
\right),
\end{equation}
is defined by:
\begin{equation}
\label{eq39}
{ }^{GVYABC}D_x^{\left( {\omega \left( x \right)} \right)} \Xi \left( x
\right)=\frac{1}{\Gamma \left( {1-\omega \left( x \right)}
\right)}\int\limits_a^x {{\rm E}_{\omega \left( x \right)} \left( {-\left(
{x-\mu } \right)^{\omega \left( x \right)}} \right)\Xi ^{\left( 1
\right)}\left( \mu \right)d\mu } ,
\end{equation}
where $a\le x$.

The plot of the different kernels of the FDs for $\omega \left( x
\right)=x-1$ is displayed in Figure 2.

\hspace{3mm}
\begin{center}
\includegraphics[width=5in]{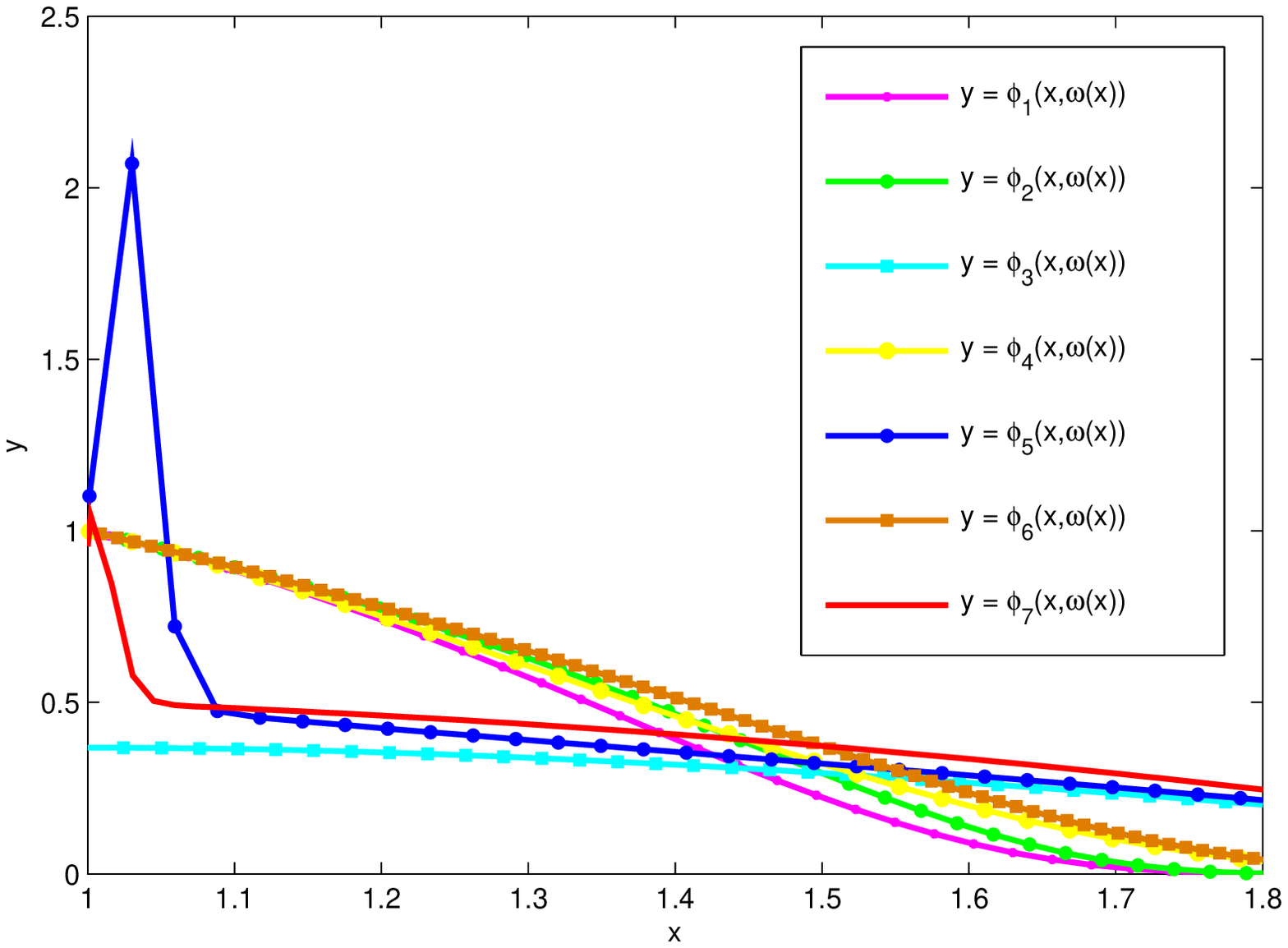}\\
{\footnotesize Figure 2. The different kernels of the FDs for $\omega \left( x
\right)=x-1$.}
\end{center}
\label{fig2}
\hspace{3mm}

The LT of the function $x^{-1-\omega \left( x \right)}\mbox{/}\Gamma \left(
{-\omega \left( x \right)} \right)$, given by Coimbra, takes the form [28]:
\begin{equation}
\label{eq40}
\widehat{L}\left[ {\frac{x^{-1-\omega \left( x \right)}}{\Gamma \left(
{-\omega \left( x \right)} \right)}} \right]=s^{\omega \left( x \right)}.
\end{equation}
The LT of the function ${\rm E}_{\omega \left( x \right)} \left( {-x^{\omega
\left( x \right)}} \right)$ is given by [14]:
\begin{equation}
\label{eq41}
\widehat{L}\left[ {{\rm E}_{\omega \left( x \right)} \left( {-\beta
x^{\omega \left( x \right)}} \right)} \right]=\frac{s^{\omega \left( x
\right)-1}}{s^{\omega \left( x \right)}+\beta }.
\end{equation}
The LTs of the FDs of variable order are as follows:
\begin{equation}
\label{eq42}
\widehat{L}\left[ {{ }^{VYCF}D_x^{\left( {\omega \left( x \right)} \right)}
\Xi \left( x \right)} \right]=\frac{1}{{1 - \omega \left( x \right)}}\sum\limits_{i = 0}^\infty  {\frac{{\left( { - \frac{{\omega \left( x \right)}}{{1 - \omega \left( x \right)}}} \right)^i \Gamma \left( {i + 1} \right)\left[ {s\hat \Xi \left( s \right) - \Xi \left( 0 \right)} \right]}}{{s^{i + 1} }}},
\end{equation}
\begin{equation}
\label{eq43}
\widehat{L}\left[ {{ }^{VYSHZD}D_x^{\left( {\omega \left( x \right)}
\right)} \Xi \left( x \right)} \right]=\frac{{\Im \left( {\omega \left( x \right)} \right)}}{{1 - \omega \left( x \right)}}\sum\limits_{i = 0}^\infty  {\frac{{\left( { - \frac{{\omega \left( x \right)}}{{1 - \omega \left( x \right)}}} \right)^i \Gamma \left( {i\omega \left( x \right) + 1} \right)\left[ {s\hat \Xi \left( s \right) - \Xi \left( 0 \right)} \right]}}{{\Gamma \left( {i + 1} \right)s^{i\omega \left( x \right) + 1} }}},
\end{equation}
\begin{equation}
\label{eq44}
\widehat{L}\left[ {{ }^{GVYSHZD}D_x^{\left( {\omega \left( x \right)}
\right)} \Xi \left( x \right)} \right]=\frac{{\Im \left( {\omega \left( x \right)} \right)}}{{1 - \omega \left( x \right)}}\sum\limits_{i = 0}^\infty  {\frac{{\left( { - 1} \right)^i \Gamma \left( {i\omega \left( x \right) + 1} \right)\left[ {s\hat \Xi \left( s \right) - \Xi \left( 0 \right)} \right]}}{{\Gamma \left( {i + 1} \right)s^{i\omega \left( x \right) + 1} }}},
\end{equation}
\begin{equation}
\label{eq45}
\widehat{L}\left[ {{ }^{VYGC}D_x^{\left( {\omega \left( x \right)} \right)}
\Xi \left( x \right)} \right]=\frac{1}{{1 - \omega \left( x \right)}}\sum\limits_{i = 0}^\infty  {\frac{{\left( { - \frac{{\omega \left( x \right)}}{{1 - \omega \left( x \right)}}} \right)^i \Gamma \left( {i + 1} \right)\left[ {s\hat \Xi \left( s \right) - \Xi \left( 0 \right)} \right]}}{{\Gamma \left( {i\omega \left( x \right) + 1} \right)s^{i + 1} }}},
\end{equation}
\begin{equation}
\label{eq46}
\widehat{L}\left[ {{ }^{GVYGC}D_x^{\left( {\omega \left( x \right)} \right)}
\Xi \left( x \right)} \right]=\frac{1}{{1 - \omega \left( x \right)}}\sum\limits_{i = 0}^\infty  {\frac{{\left( { - 1} \right)^i \Gamma \left( {i + 1} \right)\left[ {s\hat \Xi \left( s \right) - \Xi \left( 0 \right)} \right]}}{{\Gamma \left( {i\omega \left( x \right) + 1} \right)s^{i + 1} }}},
\end{equation}
\begin{equation}
\label{eq47}
\widehat{L}\left[ {{ }^{VYABC}D_x^{\left( {\omega \left( x \right)} \right)}
\Xi \left( x \right)} \right]=\frac{1}{{\Gamma \left( {1 - \omega \left( x \right)} \right)}}\frac{{s^{\omega \left( x \right) - 1} \left[ {s\hat \Xi \left( s \right) - \Xi \left( 0 \right)} \right]}}{{s^{\omega \left( x \right)}  + \frac{{\omega \left( x \right)}}{{1 - \omega \left( x \right)}}}},
\end{equation}
\begin{equation}
\label{eq48}
\widehat{L}\left[ {{ }^{GVYABC}D_x^{\left( {\omega \left( x \right)}
\right)} \Xi \left( x \right)} \right]=\frac{1}{{\Gamma \left( {1 - \omega \left( x \right)} \right)}}\frac{{s^{\omega \left( x \right) - 1} \left[ {s\hat \Xi \left( s \right) - \Xi \left( 0 \right)} \right]}}{{s^{\omega \left( x \right)}  + 1}}.
\end{equation}

\section{The constant- and variable-order anomalous relaxation problems
arising in heat transfer }

The anomalous relaxation in the sense of the constant Caputo,
Caputo-Fabrizio, Sun-Hao-Zhang- Baleanu types and variable-order FD of
Caputo type were discussed in \cite{15,23}. To present the temperature fields in
the complex media, we consider a class of anomalous relaxation equations
with the aid of the FD of constant and variable orders. The anomalous
relaxation involving the variable-order FDs are in line with the result
based on the generalized versions. Therefore, we consider the LT-type
solutions for the anomalous relaxation with the aid of the generalized FDs
of the variable order in this section.

The anomalous relaxation in the sense of the FD with respect to the
Mittag-Leffler function kernel reads
\begin{equation}
\label{eq49}
-k{ }^{YC}D_x^{\left( \omega \right)} \Xi _{YC} \left( x \right)=\Xi _{YC}
\left( x \right),
\end{equation}
with the LT of the solution
\begin{equation}
\label{eq50}
\hat {\Xi }_{YC} \left( s \right)=\frac{\frac{k\Im \left( \omega
\right)}{1-\omega }\sum\limits_{i=0}^\infty {\left( {-\frac{\omega
}{1-\omega }} \right)^i\frac{\Gamma \left( {i+1} \right)}{\Gamma \left(
{i\omega +1} \right)}\frac{1}{s^{i+1}}} }{\frac{k\Im \left( \omega
\right)}{1-\omega }\sum\limits_{i=0}^\infty {\left( {-\frac{\omega
}{1-\omega }} \right)^i\frac{\Gamma \left( {i+1} \right)}{\Gamma \left(
{i\omega +1} \right)}\frac{1}{s^i}} +1},
\end{equation}
where $k$ is a constant, and $\Xi _{YC} \left( 0 \right)=1$ is the initial
temperature field.

The anomalous relaxation involving the FD with respect to the stretched
exponential function kernel in the sense of Gaussian-like type is
\begin{equation}
\label{eq51}
-k{ }^{GYC}D_x^{\left( \omega \right)} \Xi _{GYC} \left( x \right)=\Xi
_{GYC} \left( x \right),
\end{equation}
with the LT of the solution
\begin{equation}
\label{eq52}
\hat {\Xi }_{GYC} \left( s \right)=\frac{\frac{k\left( {1+\omega ^2}
\right)}{\sqrt {\pi ^\omega \left( {1-\omega } \right)}
}\sum\limits_{i=0}^\infty {\left( {-\frac{\omega }{1-\omega }}
\right)^i\frac{\Gamma \left( {1+2\omega i} \right)}{\Gamma \left( {1+i}
\right)}\frac{1}{s^{2i\omega +1}}} }{\frac{k\left( {1+\omega ^2}
\right)}{\sqrt {\pi ^\omega \left( {1-\omega } \right)}
}\sum\limits_{i=0}^\infty {\left( {-\frac{\omega }{1-\omega }}
\right)^i\frac{\Gamma \left( {1+2\omega i} \right)}{\Gamma \left( {1+i}
\right)}\frac{1}{s^{2i\omega }}} +1},
\end{equation}
where $k$ is a constant, and $\Xi _{GYC} \left( 0 \right)=1$ is the initial
temperature field.

The anomalous relaxation involving the FD with respect to the stretched
Mittag-Leffler function kernel in the sense of Gaussian-like type can be
written as
\begin{equation}
\label{eq53}
-k{ }^{GYGC}D_x^{\left( \omega \right)} \Xi _{GYGC} \left( x \right)=\Xi
_{GYGC} \left( x \right),
\end{equation}
with the LT of the solution
\begin{equation}
\label{eq54}
\hat {\Xi }_{GYGC} \left( s \right)=\frac{\frac{k\left( {1+\omega ^2}
\right)}{\sqrt {\pi ^\omega \left( {1-\omega } \right)}
}\sum\limits_{i=0}^\infty {\left( {-\frac{\omega }{1-\omega }}
\right)^i\frac{1}{\Gamma \left( {1+i\omega } \right)}\frac{1}{s^{2i\omega
+1}}} }{\frac{k\left( {1+\omega ^2} \right)}{\sqrt {\pi ^\omega \left(
{1-\omega } \right)} }\sum\limits_{i=0}^\infty {\left( {-\frac{\omega
}{1-\omega }} \right)^i\frac{1}{\Gamma \left( {1+i\omega }
\right)}\frac{1}{s^{2i\omega }}} +1},
\end{equation}
where $k$ is a constant, and $\Xi _{GYGC} \left( 0 \right)=1$ is the initial
temperature field.

The anomalous relaxation involving the variable-order FD of Caputo-Fabrizio
type in term of the exponential kernel is:
\begin{equation}
\label{eq55}
-k{ }^{VYCF}D_x^{\left( {\omega \left( x \right)} \right)} \Xi _{VYCF}
\left( x \right)=\Xi _{VYCF} \left( x \right),
\end{equation}
with the LT of the solution
\begin{equation}
\label{eq56}
\hat {\Xi }_{VYCF} \left( s \right)=\frac{\frac{k}{1-\omega \left( x
\right)}\sum\limits_{i=0}^\infty {\left( {-\frac{\omega \left( x
\right)}{1-\omega \left( x \right)}} \right)^i\frac{\Gamma \left( {i+1}
\right)}{s^{i+1}}} }{\frac{k}{1-\omega \left( x
\right)}\sum\limits_{i=0}^\infty {\left( {-\frac{\omega \left( x
\right)}{1-\omega \left( x \right)}} \right)^i\frac{\Gamma \left( {i+1}
\right)}{s^i}} +1},
\end{equation}
where $k$ is a constant, and $\Xi _{VYCF} \left( 0 \right)=1$ is the initial
temperature field.

The anomalous relaxation in the sense of the variable-order FD of
Caputo-Fabrizio type in term of the exponential kernel is represented as:
\begin{equation}
\label{eq57}
-k{ }^{GVYSHZD}D_x^{\left( {\omega \left( x \right)} \right)} \Xi _{GVYSHZD}
\left( x \right)=\Xi _{GVYSHZD} \left( x \right),
\end{equation}
with the LT of the solution
\begin{equation}
\label{eq58}
\hat {\Xi }_{GVYSHZD} \left( s \right)=\frac{\frac{k\Im \left( {\omega
\left( x \right)} \right)}{1-\omega \left( x
\right)}\sum\limits_{i=0}^\infty {\frac{\left( {-1} \right)^i\Gamma \left(
{i\omega \left( x \right)+1} \right)}{\Gamma \left( {i+1} \right)s^{i\omega
\left( x \right)+1}}} }{\frac{k\Im \left( {\omega \left( x \right)}
\right)}{1-\omega \left( x \right)}\sum\limits_{i=0}^\infty {\frac{\left(
{-1} \right)^i\Gamma \left( {i\omega \left( x \right)+1} \right)}{\Gamma
\left( {i+1} \right)s^{i\omega \left( x \right)}}} +1},
\end{equation}
where $k$ is a constant, and $\Xi _{GYGC} \left( 0 \right)=1$ is the initial
temperature field.

The anomalous relaxation involving the FD with respect to the stretched
Mittag-Leffler function kernel in the sense of Gaussian-like type is
considered as:
\begin{equation}
\label{eq59}
-k{ }^{GVYGC}D_x^{\left( {\omega \left( x \right)} \right)} \Xi _{GVYGC}
\left( x \right)=\Xi _{GVYGC} \left( x \right),
\end{equation}
with the LT of the solution
\begin{equation}
\label{eq60}
\hat {\Xi }_{GVYGC} \left( s \right)=\frac{\frac{k}{1-\omega \left( x
\right)}\sum\limits_{i=0}^\infty {\frac{\left( {-1} \right)^i\Gamma \left(
{i+1} \right)}{\Gamma \left( {i\omega \left( x \right)+1} \right)s^{i+1}}}
}{\frac{k}{1-\omega \left( x \right)}\sum\limits_{i=0}^\infty {\frac{\left(
{-1} \right)^i\Gamma \left( {i+1} \right)}{\Gamma \left( {i\omega \left( x
\right)+1} \right)s^i}} +1},
\end{equation}
where $k$ is a constant, and $\Xi _{GVYGC} \left( 0 \right)=1$ is the
initial temperature field.

The anomalous relaxation in the sense of the variable-order FD of
Caputo-Fabrizio type in term of the exponential kernel is represented as:
\begin{equation}
\label{eq61}
-k{ }^{GVYABC}D_x^{\left( {\omega \left( x \right)} \right)} \Xi _{GVYABC}
\left( x \right)=\Xi _{GVYABC} \left( x \right),
\end{equation}
with the LT of the solution
\begin{equation}
\label{eq62}
\hat {\Xi }_{GVYABC} \left( s \right)=\frac{\frac{k}{\Gamma \left( {1-\omega
\left( x \right)} \right)}\frac{s^{\omega \left( x \right)-1}}{s^{\omega
\left( x \right)}+1}}{\frac{k}{\Gamma \left( {1-\omega \left( x \right)}
\right)}\frac{s^{\omega \left( x \right)}}{s^{\omega \left( x
\right)}+1}+1},
\end{equation}
where $k$ is a constant, and $\Xi _{GVYABC} \left( 0 \right)=1$ is the
initial temperature field.

\section{Conclusions}
In this work, we suggested a class of the FDs of constant and variable
orders for the first time. The plots of the kernel functions of the
different patterns in were discussed in detail. The LT-type solutions for
the anomalous relaxations involving the FDs of constant and variable orders
were also given. The proposed formulas are useful to open up the new
prospects of describing the fractional-order heat-transfer equations in the
complex media.



\begin{thebibliography}{11}


\bibitem{1}
West, B. J., \textit{Fractional Calculus View of Complexity: Tomorrow's Science}, CRC Press, Boca Raton, 2015

\bibitem{2}
Mainardi, F.,~\textit{Fractional Calculus and Waves in Linear Viscoelasticity: an Introduction to Mathematical Models} World Scientific, 2010

\bibitem{3}
Herrmann, R.,~\textit{Fractional Calculus: An Introduction for Physicists}, World Scientific, 2014

\bibitem{4}
He, J. H., A New Fractal Derivation, \textit{Thermal Science},~15(2011), Suppl. 1, pp. 145--147

\bibitem{5}
Yang, X. J., \textit{et al.,} \textit{Local Fractional Integral Transforms and Their Applications}, Academic Press, 2015

\bibitem{6}
Yang, X. J., \textit{et al.,} On Local Factional Operators View of Computational Complexity£º Diffusion and Relaxation Defined on Cantor Sets, \textit{ Thermal Science},~20 (2016), Suppl. 3, pp. 145--147

\bibitem{7}
Yang, X. J., \textit{et al.,} On a Fractal LC-electric Circuit Modeled by Local Fractional Calculus, \textit{Communications in Nonlinear Science and Numerical Simulation}, 47 (2017), pp.200--206

\bibitem{8}
Mainardi, F., Gorenflo, R., Time-fractional Derivatives in Relaxation Processes: a Tutorial Survey, \textit{~Fractional Calculus and Applied Analysis}~10(2007), 3, pp.269--308

\bibitem{9}
Mainardi, F., Spada, G., Creep, relaxation and viscosity properties for basic fractional models in rheology,~\textit{The European Physical Journal Special Topics},~193(2011), 1, pp.133--160

\bibitem{10}
Kilbas, A. A., \textit{et al.,} \textit{Theory and Applications of Fractional Differential Equations}, Elsevier, Amsterdam, The Netherlands, 2006

\bibitem{11}
Samko, S. G., Ross, B., Integration and Differentiation to a Variable Fractional Order, \textit{Integral Transforms and Special Functions},~1 (1993), 4, pp.277--300

\bibitem{12}
Kilbas, A. A., \textit{et al.,} \textit{Fractional Integral and Derivatives: Theory and Applications}~Gordon and Breach, Switzerland, 1993

\bibitem{13}
Lorenzo, C. F., Hartley, T. T., Variable Order and Distributed Order Fractional Operators, \textit{Nonlinear dynamics}, 29(2002), 1-4, pp.57--98

\bibitem{14}
Coimbra, C. F., Mechanics with Variable Order Differential Operators, \textit{Annalen der Physik}, 12 (2003), 11-12, pp.692--703

\bibitem{15}
Ramirez, L. E. S., Coimbra, C. F. M., A Variable Order Constitutive Relation for Viscoelasticity, \textit{Annalen der Physik},~16(2007), 78, pp.543--552

\bibitem{16}
Bhrawy, A. H. Zaky M. A., Numerical Algorithm for the Variable-order Caputo Fractional Functional Differential Equation, \textit{Nonlinear Dynamics}, 85(2016), 3, pp 1815--1823

\bibitem{17}
Caputo, M., Fabrizio, M. A., New Definition of Fractional Derivative without Singular Kernel,\textit{ Progress in Fractional Differentiation and Applications}, 1 (2015), 2, pp.73--85

\bibitem{18}
Alkahtani, B. S. T., Atangana, A., Controlling the Wave Movement on the Surface of Shallow Water with the Caputo--Fabrizio Derivative with Fractional Order, \textit{Chaos, Solitons {\&} Fractals}, 89 (2016), pp.539--546

\bibitem{19}
Losada, J., Nieto, J. J., Properties of a New Fractional Derivative without Singular Kernel, \textit{Progress in Fractional Differentiation and Applications}, 1 (2015), 2, pp.87--92

\bibitem{20}
Caputo, M., Fabrizio, M., Applications of New Time and Spatial Fractional Derivatives with Exponential Kernels, \textit{Progress in Fractional Differentiation and Applications}, 2 (2016), 2, pp.1--11

\bibitem{21}
Yang, X. J., \textit{et al}., A New Fractional Derivative without Singular Kernel: Application to the Modelling of the Steady Heat Flow, \textit{Thermal Science}, 20(2016), 2, pp.753--756

\bibitem{22}
Yang, A. M., \textit{et al.,} On Steady Heat Flow Problem Involving Yang-Srivastava-Machado Fractional Derivative Without Singular Kernel, \textit{Thermal Science}, 20 (2016), Suppl. 3, pp. 717--723

\bibitem{23}
Sun, H.,~\textit{et al.,} Relaxation and Diffusion Models with Non-singular Kernels \textit{Physica A}, 2016, DOI: 10.1016/j.physa.2016.10.066

\bibitem{24}
Atangana, A., Dumitru, B., New Fractional Derivatives with Non-local and Non-singular Kernel: Theory and Application to Heat Transfer Model, \textit{Thermal Science}, 20 (2016), 2, pp.763--769


\end{thebibliography}
\end{document}